\documentclass[twocolumn]{aastex7}

\usepackage{amsmath}
\usepackage{amssymb}
\usepackage{amsfonts}
\usepackage{graphicx}
\usepackage{multirow}
\usepackage{bm}
\usepackage{hyperref}
\hypersetup{
  colorlinks=true,
  linkcolor=  blue,
  citecolor=  blue,
  urlcolor=   blue,
  anchorcolor=blue,
  linkcolor=  blue
  }
\usepackage[normalem]{ulem}
\usepackage{epsfig}
\usepackage{verbatim}
\usepackage{color}
\usepackage{mdframed}
\usepackage{soul} 
\usepackage{braket}
\usepackage{enumitem}
\usepackage[percent]{overpic}
\usepackage{float}
\usepackage{array,booktabs,tabularx}
\renewcommand{\arraystretch}{1.6}
\usepackage{soul}
\include{eps2pdf}
\usepackage{fontawesome}
\usepackage{sidecap}
\usepackage{multirow}
\usepackage[flushmargin,bottom]{footmisc}
\usepackage{longtable}
\usepackage[normalem]{ulem}

\shorttitle{Correcting for interloper contamination in the power spectrum}
\shortauthors{Cagliari et al.}

\begin{document}

\title{Correcting for interloper contamination in the power spectrum with neural networks}

\author[]{Marina S. Cagliari}
\affiliation{Laboratoire d’Annecy de Physique Theorique (LAPTh), CNRS/USMB, 99 Chemin de Bellevue BP110 - Annecy - \\ F-74941 - ANNECY CEDEX - FRANCE}
\email[show]{marina.cagliari@lapth.cnrs.fr\\}  

\author[]{Azadeh Moradinezhad Dizgah}
\affiliation{Laboratoire d’Annecy de Physique Theorique (LAPTh), CNRS/USMB, 99 Chemin de Bellevue BP110 - Annecy - \\ F-74941 - ANNECY CEDEX - FRANCE}
\altaffiliation{azadeh.moradinezhad@lapth.cnrs.fr}
\email[]{azadeh.moradinezhad@lapth.cnrs.fr}

\author[]{Francisco Villaescusa-Navarro}
\affiliation{Center for Computational Astrophysics, Flatiron Institute, 162 5th Avenue, New York, NY 10010, USA}
\affiliation{Department of Astrophysical Sciences, Princeton University, Peyton Hall, Princeton, NJ 08544-0010, USA}
\email[]{fvillaescusa@flatironinstitute.org}

%% Use the \collaboration command to identify collaborations. This command
%% takes an optional argument that is either a number or the word "all"
%% which tells the compiler how many of the authors above the command to
%% show. For example "\collaboration[all]{(DELVE Collaboration)}" wil include
%% all the authors above this command.
%%
%% Mark off the abstract in the ``abstract'' environment. 

\begin{abstract}
Modern slitless spectroscopic surveys, such as \textit{Euclid} and the Nancy Grace Roman Space Telescope, collect vast numbers of galaxy spectra but suffer from low signal-to-noise ratios. This often leads to incorrect redshift assignments when relying on a single emission line, due to noise spikes or contamination from non-target emission lines, commonly referred to as {\it redshift interlopers}. We propose a machine learning approach to correct the impact of interlopers at the level of measured summary statistics, focusing on the power spectrum monopole and line interlopers as a proof of concept. To model interloper effects, we use halo catalogs from the \textsc{Quijote} simulation suite as proxies for galaxies, displacing a fraction of halos by the distance corresponding to the redshift offset between target and interloper galaxies. This yields contaminated catalogs with varying interloper fractions across a wide range of cosmologies from the \textsc{Quijote} suite. We train a neural network on power spectrum monopole measurements, alone or combined with the bispectrum monopole, from contaminated mocks to estimate the interloper fraction and reconstruct the cleaned power spectrum. We evaluate performance in two settings: one with fixed cosmology and another where cosmological parameters vary under broad priors. In the fixed case, the network recovers the interloper fraction and corrects the power spectrum to better than $1\%$ accuracy. When cosmology varies, performance degrades, but incorporating bispectrum information significantly improves results, reducing the interloper fraction error by $40$--$60\%$. Catastrophic failures occur mainly at extreme cosmological values, which are unlikely in real data. We also study the method’s performance as a function of the size of the training set and find that optimal strategies depend on the correlation between target and interloper samples: bispectrum information aids performance when target and interloper galaxies are uncorrelated, while tighter priors are more effective when the two are strongly correlated.
\end{abstract}

%% Keywords should appear after the \end{abstract} command. 
%% The AAS Journals now uses Unified Astronomy Thesaurus (UAT) concepts:
%% https://astrothesaurus.org
%% You will be asked to selected these concepts during the submission process
%% but this old "keyword" functionality is maintained in case authors want
%% to include these concepts in their preprints.
%%
%% You can use the \uat command to link your UAT concepts back its source.

%% From the front matter, we move on to the body of the paper.

%%%%%%%%%%%%%%%%%%%%%%%%
\section{Introduction\label{sec:introduction}}
%%%%%%%%%%%%%%%%%%%%%%%%

Space-based Stage-IV galaxy surveys, such as the ESA's \textit{Euclid} satellite\footnote{\url{https://www.euclid-ec.org}}~\citep{Euclid:2024yrr} and the NASA's Nancy Grace Roman Space Telescope\footnote{\url{https://roman.gsfc.nasa.gov}}~\citep{Wang:2021oec}, use slitless spectroscopy to measure the galaxy spectra. These missions target emission line galaxies (ELGs), determining their redshifts using one or more prominent emission lines. The advantage of slitless spectroscopy is its ability to simultaneously acquire spectra for a large number of galaxies over wide areas of the sky, enabling coverage of vast cosmological volumes. However, this comes at the cost of a lower emission line signal-to-noise ratio compared to ground-based experiments such as the Dark Energy Spectroscopic Instrument\footnote{\url{https://www.desi.lbl.gov/}}~\citep{desi}.
 
Determination of redshifts in slitless surveys faces two main challenges. The first involves intense noise spikes in the spectra, which can be erroneously classified as emission lines. This effect stochastically adds objects with incorrect redshifts to the galaxy sample and is commonly referred to as \emph{noise interlopers}. The second issue arises from the presence of additional spectral emission lines beyond the target line. Due to limited spectral resolution, these spurious lines can not be reliably distinguished from the target line, leading to the inclusion of galaxies with incorrect redshifts. These are commonly referred to as \emph{line interlopers}.

The Euclid Wide Survey \citep{Euclid:2021icp} is a prime example of a survey where the final sample will include target galaxies, line interlopers, and noise interlopers. The reason of this being that \textit{Euclid} does not have a spectroscopic resolution high enough to separate the H$\alpha$--[N$_{\rm II}$] complex and distinguish it from the [O$_{\rm III}$] doublet of higher redshift objects; hence, [O$_{\rm III}$] emitters may be misclassified as H$\alpha$ galaxies introducing contamination in the target sample. Similarly, the [S$_{\rm III}$] line in low redshift emitters can be interpreted as an H$\alpha$ emission. In the case of Roman, if only one emission line were to be used to determine the galaxy redshifts, H$\beta$ emitters may be taken for [O$_{\rm III}$] galaxies, which now are the targets. The only difference is that the contaminants now come from the same redshift as the targets.

If not properly accounted for, the presence of noise and line interlopers can severely bias the constraints on cosmological parameters derived from galaxy clustering statistics  \citep{Pullen:2015yba,Addison:2018xmc,Massara:2020abv,Gong:2021qlj}. To mitigate this, one must either identify and remove contaminant galaxies at the catalog level using target selection schemes \citep{Euclid:2024rko}, or model their impact directly in the summary statistics used for inference. Several previous works have investigated how to model the interloper effect in different observables, such as the the baryon acoustic oscillations \citep[BAO;][]{Addison:2018xmc,Massara:2020abv,Nguyen:2023xel,Foroozan:2022lwh}, the two-point correlation function \citep{Farrow:2021trw,Risso-prep}, and the power spectrum \citep{Addison:2018xmc,GrasshornGebhardt:2018cho,Lee-prep}, to correct for biases in cosmological inference. These methods typically require the interloper fraction as either an input or a nuisance parameter in the analysis. In the former case, the interloper fraction can be estimated from deep spectroscopic observations, though such data may be limited in availability.  

Considering line interlopers, \cite{Massara:2023ull} proposed a machine learning approach to predict the interloper fraction in galaxy catalogs using graph neural networks, learning its posterior distribution while marginalizing over cosmological parameters. Motivated by these results, in this work, we explore the possibility of training deep neural networks to infer the interloper fraction directly from measured summary statistics, rather than from the catalogs, and to correct their impact on the statistics.

We develop a moment neural network that takes interloper-contaminated summary statistics as input and outputs both the mean and variance of the interloper fraction, along with a correction to remove the contamination from the power spectrum monopole. By learning to model the uncertainty in the interloper fraction, the network provides a more robust correction than point estimates alone. A key advantage of our approach is that, by correcting the observed summary statistics directly, it removes the need to explicitly model interlopers in the theoretical predictions used for cosmological inference. Furthermore, in contrast to the catalog-level method proposed by \citet{Massara:2023ull}, which uses graph neural networks and is limited to small datasets, our framework scales efficiently and can be applied to large-volume surveys.

To train and test our machine learning approach, we generate simulated snapshots based on the \textsc{Quijote} suite \citep{Villaescusa-Navarro:2019bje}, introducing line interlopers to mimic realistic survey conditions. We consider two scenarios: interlopers originating from a different redshift than the target sample, as expected for Euclid-like observations, and interlopers at a similar redshift to the targets, reflecting a Roman-like case. We construct these contaminated simulations under two configurations: one with fixed cosmology and varying interloper fraction, and another where both cosmological parameters and the interloper fraction vary. To train the neural network, we test two types of input features: first, the interloper-contaminated power spectrum monopole alone, and then the combination of power spectrum and bispectrum monopoles. In the varying cosmology scenario, we provide the network with cosmological parameter priors and investigate how the prior width affects performance. We also explore how the network's performance scales with the size of the training set, aiming to determine the minimum number of simulations required for efficient learning.

The simulations prepared for this work are available as part of the \textsc{Quijote} suite,\footnote{\url{https://quijote-simulations.readthedocs.io/en/latest/interlopers.html}} and the code is public.\footnote{\url{https://github.com/mcagliari/NoInterNet}}

The remainder of this paper is structured as follows. In section~\ref{sec:simulations}, we describe the simulations used to model interlopers and the summary statistics we employ in our analysis. Section~\ref{sec:NN} presents our neural network approach, detailing the architecture, input and output features, and the performance metrics used. We report our results in section~\ref{sec:results}, and conclude in section~\ref{sec:conclusions}.

%%%%%%%%%%%%%%%%%%%%%%%%
\section{Simulation Setup}
\label{sec:simulations}
%%%%%%%%%%%%%%%%%%%%%%%%

In this section, we describe the data generation pipeline used to train and evaluate our machine learning framework. We begin with an overview of the \textsc{Quijote} simulation suite, which provides the cosmological $N$-body simulations used as our baseline. We then detail how we inject interloper contamination into these simulations to mimic both \textit{Euclid}-like and Roman-like scenarios. Finally, we describe how we measure the power spectrum and bispectrum monopoles from the simulated data, which serve as inputs to our neural network.

%----------------------
\subsection{\textsc{Quijote} Simulation Suite}
\label{sec:quijote}
%----------------------

The \textsc{Quijote} simulation suite \citep{Villaescusa-Navarro:2019bje} consists of a large number of $N$-body simulations run with the \texttt{GADGET-III} TreePM+SPH code \citep{Springel:2005mi}, specifically designed to support machine learning applications in cosmology. Each (mid-resolution) simulation evolves $512^3$ particles in a periodic box of $L_{\rm box} = 1 \, \mathrm{Gpc}/h$ on the side, from redshift $z = 127$ to $z = 0$. Initial conditions are generated using second-order Lagrangian Perturbation Theory (2LPT) via the \texttt{2LPTIC} code.\footnote{\url{https://cosmo.nyu.edu/roman/2LPT/}} The suite includes five snapshots of the dark matter particle at redshifts $z = \{0, 0.5, 1, 2, 3\}$, along with two halo catalogs per snapshot. Halos are identified using either the Friend-of-Friend \citep[FoF;][]{FoF} or the \textsc{Rockstar} \citep{rockstar} halo finding algorithms.

We utilize the FoF halo catalogs at redshifts $z = \{1, 2\}$, and consider two sets of \textsc{Quijote} simulations. First, to test our algorithm in a controlled setting, we consider the fixed cosmology case using a subset of $N_{\rm fid} = 1000$ simulations from the fiducial dataset. These are independent realizations of a flat $\Lambda$CDM cosmology with $\Omega_m = 0.3175$, $\Omega_b = 0.049$, $h = 0.6711$, $n_s = 0.9624$, and $\sigma_8 = 0.834$. Second, to evaluate the performance of our approach under cosmological uncertainty, reflecting a more realistic application to survey data, we use the \textsc{Quijote} Big Sobol Sequence \citep[BSQ;][]{Bairagi:2025sux}. This dataset includes $N_{\rm BSQ} = 2^{15}$ flat $\Lambda$CDM cosmologies, with parameters sampled from a Sobol sequence within the prior ranges: $\Omega_m \in [0.1, 0.5]$, $\Omega_b \in [0.02, 0.08]$, $h \in [0.5, 0.9]$, $n_s \in [0.8, 1.2]$, and $\sigma_8 \in [0.6, 1.0]$. In both the fiducial and BSQ simulations, the total neutrino mass is set to zero.

%----------------------
\subsection{Simulating Interloper Contamination}
\label{sec:interlop-sim}
%----------------------

Line interlopers are galaxies that are at the true redshift $z_{\rm t}$, but are considered to be at the incorrectly-determined redshift $z_{\rm f}$. This incorrect redshift estimation leads to a shift in radial comoving distance, $\chi$, of interloper galaxies:
\begin{equation}
    \Delta \chi = \chi(z_{\rm f}) - \chi(z_{\rm t}) \, ,
    \label{eq:true_displacement}
\end{equation}
with the incorrect redshift given by 
\begin{equation}
    z_{\rm f} = \frac{\lambda_{\rm t} \, (1 + z_{\rm t}) - \lambda_{\rm f}}{\lambda_{\rm f}} \, ,
    \label{eq:obs-z}
\end{equation}
where $\lambda_{\rm t}$ is the true rest frame wavelength of the observed line (e.g., [O$_{\rm III}$] or [S$_{\rm III}$]) and $\lambda_{\rm f}$ the rest frame wavelength of the target line (e.g., H$\alpha$).

Depending on the spectrometer wavelength range and the target line, interlopers may come from a population with a redshift close to the target sample ([O$_{\rm III}$]--H$\beta$ interlopers for the Roman Space Telescope) or from a very different redshift ([O$_{\rm III}$]/[S$_{\rm III}$]--H$\alpha$ interlopers for \textit{Euclid}). In the first case, we can simplify equation~\eqref{eq:true_displacement} \citep{Massara:2023ull}, 
\begin{equation}
    \Delta \chi \approx \frac{c \, (1 + z_{\rm f})}{H(z_{\rm f})} \, \left( 1 - \frac{\lambda_{\rm t}}{\lambda_{\rm f}} \right) \, ,
    \label{eq:inbox-displacement}
\end{equation}
as $z_{\rm f} \approx z_{\rm t}$ and the interloper and target population are strongly correlated. On the other hand, when interlopers come from a different redshift, the interloper galaxies will bring a clustering contribution to the power spectrum that is not (strongly) correlated to the target clustering and we have to determine their shift along the line-of-sight with equations~\eqref{eq:true_displacement} and \eqref{eq:obs-z}. We will discuss in more detail the effect of the two types of line interlopers on the observed clustering in section~\ref{sec:interlop-sim}.

To accurately model the impact of interlopers, constructing lightcone simulations is necessary, as it accounts for contamination of the target galaxy sample from sources at different redshifts. However, to test our method in a more controlled environment, we start from the simplified case of simulated snapshots at fixed redshift. Above, we discussed that line interlopers come from galaxy populations that can either be close in redshift to the targets or from different redshifts (higher or lower redshifts depending on the emission line). We use two different pipelines to simulate the close and far redshift interlopers. Throughout this work, we refer to interlopers close in redshift to the target galaxies as \textit{inbox} interlopers, and those from distant redshifts as \textit{outbox} interlopers.

To simulate inbox interlopers, we follow the same strategy described in \cite{Massara:2023ull}. Given the box at $z=1$ with comoving coordinates $(\tilde{x},\tilde{y}, \tilde{z})$, we randomly select a fraction $f$ of objects and displace them by $\Delta \tilde{z} = 90 \, {\rm Mpc}/h$ along the $\tilde{z}$-axis, which we consider our radial coordinate, applying boundary conditions. We then apply the redshift space distortion (RSD) on shifted halos. In the outbox case, we use two snapshots: one at $z=1$, containing the target galaxies, and another at $z=2$, from which the interlopers are selected. First, we rotate the box at $z=2$ to remove the spatial correlation between the two snapshots, then we shift the second box along the $\tilde{z}$-axis ($\Delta \tilde{z}_{\rm box} = 1265 \, {\rm Mpc}/h$) and we displace a fraction $f_{\rm out}$ of halos at $z=2$ by $\Delta \tilde{z} = -1320 \, {\rm Mpc}/h$, where
\begin{equation}
    f_{\rm out} = \frac{f \, N_t}{(1 - f) \, N_o} \, ,
    \label{eq:out-frac}
\end{equation}
with $f$ the actual fraction of interlopers in the final sample (as for the inbox case), $N_t$ the number of targets, hence the number of halos at $z=1$, and $N_o$ is the total number of halos at $z=2$. We remark that the use of $f_{\rm out}$ to select the number of objects shifted from the interloper to the target box is necessary to ensure that the fractions of interlopers in the observed sample are consistent between the outbox and inbox cases. Then we apply RDS taking into account the difference in the true redshifts of the targets and interlopers. In the case of fixed cosmology, starting from $N_{\rm fid} = 1000$ simulations, we build a dataset of $2000$ contaminated simulations with $100$ fractions $f$ sampled from a Latin hypercube in the range of $[0.01,0.11]$. Therefore, in this dataset, each \textsc{Quijote} simulation has two fraction realizations and multiple simulations share the same interloper fraction.

When cosmology varies with the fraction of interlopers, to mimic the assumption of a fiducial cosmology when we measure the summary statistics, we distort the box comoving coordinates, $(\tilde{x}, \tilde{y}, \tilde{z})$, as follows
\begin{equation}
    x' = \frac{\tilde{x}}{a_{\perp}}\, , \quad y' = \frac{\tilde{y}}{a_{\perp}}\, , \quad z' = \frac{\tilde{z}}{a_{\parallel}}\, , 
    \label{eq:AP-distrition}
\end{equation}
where 
\begin{equation}
    a_\perp = \frac{H(z)}{H_{\rm fid}(z)} \, , \quad a_\parallel = \frac{D_{A, {\rm fid}}(z)}{D_A(z)} \, , 
    \label{eq:AP-params}
\end{equation}
with $H(z)$ being the Hubble parameter and $D_A(z)$ the angular diameter distance. The subscript `fid' indicates the value of the parameters in the fiducial cosmology, which we assume to be the cosmology of the \textsc{Quijote} fiducial dataset. Finally, the parameters are estimated at the redshift of the target sample, both for inbox and outbox interlopers. In the varying cosmology case, we use the BSQ \textsc{Quijote} set including $2^{15}$ cosmologies. In this case, we construct a Sobol sequence of $2^{15}$ fractions, $f \in [0.01,0.11]$, to build the interloper-contaminated simulations.

%---------------------
\subsection{Summary Statistics} 
\label{sec:summary-stats}
%---------------------
%
\begin{figure*}
\centering    \includegraphics[width=0.8\textwidth]{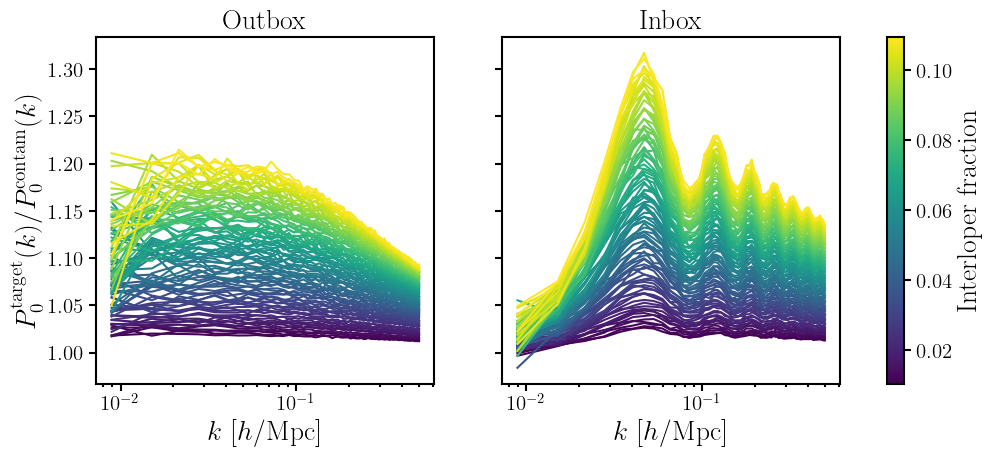}
    \caption{Ratio between target and contaminated halo power spectrum monopoles in redshift space for fixed cosmology. Lines are color-coded according to the interloper fraction. \emph{Left}: outbox interlopers. \emph{Right}: inbox interlopers.}
    \label{fig:interlop-effect}
\end{figure*}
\begin{figure*}
    \centering
    \includegraphics[width=0.8\textwidth]{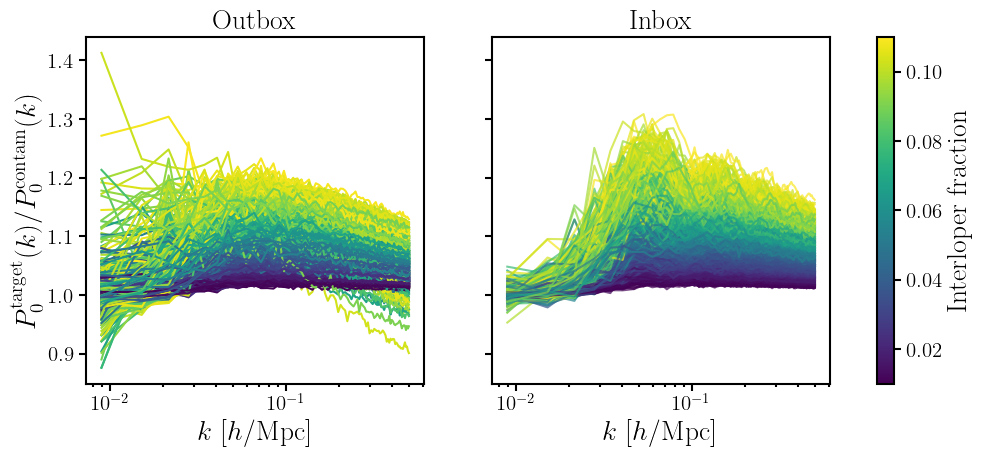}
    \caption{Same as figure~\ref{fig:interlop-effect}, but for the case where both cosmology and interloper fraction are varied.} \vspace{0.1in}
    \label{fig:interlop-effect-BSQ}
\end{figure*}
For each box we measure the power spectrum of the target sample, $P^{\rm target}(k)$, which for the outbox interlopers corresponds to the original halos of the $z=1$ box and for the inbox interlopers to the halos in the box that were not selected as interlopers and shifted. For the interloper-contaminated sample, we measure both the power spectrum,  $P^{\rm contam}(k)$, and bispectrum, $B^{\rm contam}(k_1, k_2, k_3)$, to test to what extend addition of the bispectrum information improves the performance of the method in comparison to only using the power spectrum to clean the power spectrum. 

We measure the power spectra with \texttt{Pylians3} \citep{Pylians} on a grid of $N_{\rm g} = 512^3$, while for the bispectra we use \texttt{pyspectrum} \citep{2015PhRvD..92h3532S,2020JCAP...03..040H} on a $N_{\rm g} = 256^3$ grid. We use the binning of  $\Delta k = k_f$ for the power spectrum and  $\Delta k = 6 k_f$ for the bispectrum, where $k_f = 6.28 \times 10^{-3} \, h/{\rm Mpc}$ is the fundamental mode of the simulation box. The choice of a wider binning for the bispectrum is primarily made to reduce the scatter in the measurements.  We limit the analysis for both the power spectra and bispectra to scales below the cutoff of $k_{\rm max} = 0.5 \, h/{\rm Mpc}$.

In figures~\ref{fig:interlop-effect} and~\ref{fig:interlop-effect-BSQ} we present the effect of the interlopers on the monopole of the power spectrum for the fixed and varying cosmology cases, respectively. We quantify the impact of the interlopers as the ratio between the target and contaminated power spectrum monopole, which we plot as a function of interloper fraction. The effect is more clear in figure~\ref{fig:interlop-effect}, where only the interloper fraction varies over the realizations. In this case, we see that the outbox interlopers have a scale-dependent effect, changing primarily the amplitude of the power spectrum. The inbox interlopers, on the other hand, have a strong impact on the BAO in addition to modifying the overall amplitude. These dependencies are less apparent when both the interloper fraction and cosmological parameters are allowed to vary simultaneously (see figure~\ref{fig:interlop-effect-BSQ}), due to degeneracies between them. While changes in amplitude, driven by increased number density, are still observable, the dependence on the interloper fraction is no longer unambiguous. In the case of inbox interlopers, the impact on the BAO feature also becomes less pronounced.

%%%%%%%%%%%%%%%%%%%%%%%%
\section{Neural Network Design and Evaluation}
\label{sec:NN}
%%%%%%%%%%%%%%%%%%%%%%%%

In this section, we describe the neural network architecture, its inputs and outputs, how we pre- and post-process the data, and the metrics we use to evaluate the algorithm's performance.

\begin{table*}[t]
\caption{\label{tab:architecture}Specifications of the employed neural network architecture in this study.}
\begin{ruledtabular}
\begin{tabular}{cccccccc}
Cosmology & Input & Input size & $n_{\rm in}$ & $n_{\rm min}$ & $n_{\rm out}$ & $n_b$ & Patience\\ \midrule
\multirow{2}{*}{Fixed} & $P_0^{\rm contam}$ & 79 & 64 & 32 & 64 & 64 & 500 \\ \cmidrule(l){2-8}
& \multirow{1}{*}{$\{P_0^{\rm contam}, B_0^{\rm contam}\}$} & 373 & 256 & 16 & 64  & 8 & 200 \\ \midrule
\multirow{2}{*}{Varying} & $\{P_0^{\rm contam}, \Omega_{\rm m}, \Omega_{\rm b}, h, n_{\rm s}, \sigma_8\}$ & 84 & \multirow{2}{*}{64} & \multirow{2}{*}{16} & \multirow{2}{*}{64} & \multirow{2}{*}{64} & \multirow{2}{*}{500} \\ \cmidrule(l){2-3}
& \multirow{1}{*}{$\{P_0^{\rm contam}, B_0^{\rm contam}, \Omega_{\rm m}, \Omega_{\rm b}, h, n_{\rm s}, \sigma_8\}$} & \multirow{1}{*}{378} & & & &  
\end{tabular}
\end{ruledtabular}\vspace{0.16in}
\end{table*}

%----------------------
\subsection{Network Description}
\label{sec:nointernet}
%----------------------

The baseline architecture of our model is a dense neural network that takes the contaminated summary statistic as input and outputs a set of predicted quantities: the correction to be applied to the power spectrum monopole to remove interloper contamination, the interloper fraction, and the associated uncertainties of these quantities.
A network with this output configuration is also referred to as a moment network \citep{Jeffrey:2020itg,CAMELS:2021raw}, as it is trained to output the first and second moments of the output probability distribution, hence its mean and standard deviation. The output vector is defined as
\begin{equation} \mathbf{y} = \{\mathbf{y}_{\rm c}, y_f, \bm{\sigma}_{\rm c}, \sigma_{f}\},  \label{eq:output} 
\end{equation}
where $\mathbf{y}_{\rm c}$ is the predicted first moment of the correction to the power spectrum monopole, and its true value is given by
\begin{equation} 
    \widehat{\mathbf{y}}_{\rm c} = \frac{P_0^{\rm target}(k)}{P_0^{\rm contam}(k)} , ,
    \label{eq:ycorr} 
\end{equation}
with $P_0^{\rm target}(k)$ denoting the target power spectrum monopole and $P_0^{\rm contam}(k)$ the observed, interloper-contaminated monopole. The scalar $y_f$ corresponds to the predicted interloper fraction, with true value $f$, and $\bm{\sigma}_{\rm c}$ and $\sigma_f$ represent the predicted second moments of the monopole correction and interloper fraction, respectively. In all configurations, the network output vector has a fixed dimension of 160. Boldface symbols indicate array-valued outputs.

As input, the network can take either the contaminated power spectrum monopole, $P_0^{\rm contam}(k)$, or the concatenation of the contaminated power spectrum and bispectrum monopoles, ${P_0^{\rm contam}(k), B_0^{\rm contam}(k_1, k_2, k_3)}$. When cosmological parameters are allowed to vary, we also provide the network with prior information on cosmology. Specifically, we supply the five cosmological parameters, ${\Omega_{\rm m}, \Omega_{\rm b}, h, n_{\rm s}, \sigma_8}$, drawn from a Gaussian distribution centered on the true values used in the simulation, with a standard deviation equal to $N$ times the corresponding Planck uncertainties \citep{2020A&A...641A...9P}. The normalization applied to all network inputs and outputs is described in section~\ref{sec:proc-metrics}.

The neural network has a very simple architecture composed of a stack of dense layers with a LeakyReLU as activation function \citep{Maas2013RectifierNI}, except for the output layer that has no activation. First, it compresses the input into $n_{\rm in}$ neurons and then down to some $n_{\rm min}$; second, it decompresses it to $n_{\rm out}$ neurons and then into the output dimension. During each compression step (except for going from the network input to $n_{\rm in}$), the number of neurons in the layer is halved, while in the decompression (except for going from $n_{\rm out}$ to the network output) they are doubled. The architecture is the same for the two types of interlopers, but we change it depending on the inputs. We summarize the main features of the networks with different configurations in Table~\ref{tab:architecture}.

To train the network, we use the adaptive moment estimation optimizer \citep[Adam;][]{kingma2014adam} with initial learning rate $l_r = 0.001$. The loss function for one batch is \citep{Jeffrey:2020itg,CAMELS:2021raw},
\allowdisplaybreaks{
\begin{align}
        L(\mathbf{y}, \mathbf{\widehat{y}}) & = \log \left( \frac{1}{n_b} \sum_{i = 1}^{n_b} \left( \sum_{j = 1}^{n_k} \left( \widehat{y}_{{\rm c}, ij} - y_{{\rm c}, ij} \right)^2 \right) \right) \notag \\
        & + \log \left( \frac{1}{n_b} \sum_{i = 1}^{n_b} \left( f_i - y_{f, i} \right)^2 \right) \notag \\
        & + \log \left( \frac{1}{n_b} \sum_{i = 1}^{n_b} \left( \sum_{j = 1}^{n_k} \left ( \left( \widehat{y}_{{\rm c}, ij} - y_{{\rm c}, ij} \right)^2 - \sigma_{{\rm c}, ij}^2 \right)^2 \right) \right) \notag \\
        & + \log \left( \frac{1}{n_b} \sum_{i = 1}^{n_b} \left( \left( f_i - y_{f, i} \right)^2 - \sigma_{f, i}^2 \right)^2 \right) \, ,
    \label{eq:loss}
\end{align}}
where $\mathbf{\widehat{y}} = \{\mathbf{\widehat{y}}_{\rm c}, f\}$, $n_b$ is the batch size, and $n_k$ is the number of $k$-bins in the power spectrum. Then the loss is averaged over all the batches. The first two terms in equation~\eqref{eq:loss} are the logarithms of the standard mean squared error loss function, which minimizes the mean or first moment of the distribution, while the second two terms minimize the standard deviation of the output, enabling the network to provide a statistical uncertainty related to its prediction. Finally, we train the network up to $2000$ epochs, but we also implemented an early stopping mechanism, whose patience we report in Table~\ref{tab:architecture}.

\begin{table*}
\caption{\label{tab:metrics-fixed} The MSE and $\chi_{\rm red}^2$ values for the fixed cosmology case. The $\chi_{\rm red}^2$ values marked by * are computed removing the points with predicted uncertainty $<10^{-3}$.}
\begin{ruledtabular}
\begin{tabular}{ccccc}
Input & \multicolumn{2}{c}{Outbox} & \multicolumn{2}{c}{Inbox} \\ \midrule
 & MSE & $\chi_{\rm red}^2$ & MSE & $\chi_{\rm red}^2$ \\ \midrule
Contam. $P_0$ & $8.86 \times 10^{-6}$ & 1.31 & $1.30 \times 10^{-5}$ & 1.50 \\ \midrule
Contam. $P_0 + B_0$ & $6.32 \times 10^{-6}$ & 1.48 & $9.91 \times 10^{-6}$ & 2.42*
\end{tabular}
\end{ruledtabular}\vspace{0.1in}
\end{table*}

%----------------------
\subsection{Data Processing and Performance Metrics}
\label{sec:proc-metrics}
%----------------------

As mentioned in section~\ref{sec:interlop-sim}, we train the network using two datasets: in the first one, only the interloper fraction is varied, and values of cosmological parameters are fixed to the fiducial ones, while in the second set, both cosmological parameters and interloper fractions are varied. For both datasets, we adopt the same data split of $75$, $15$, and $10\%$ for training, validation, and test sets, respectively. 

We pre-process the inputs and network labels independently of the network configuration (interloper type, variation of cosmology, and network inputs). We normalize the input to lie in the interval of $[0,1]$ and label vectors element-wise. Given the vector $\mathbf{a}$ with elements $a_i$ and part of the dataset $\mathbf{A}$, we pre-process it as follows
\begin{equation}
    a_i^{\rm n} = \frac{a_i - \min_j \left( A_{ij} \right)}{\max_j \left( A_{ij} \right) - \min_j \left( A_{ij} \right)} \, ,
\end{equation}
where $j$ runs over the dataset elements.

Finally, we define the metrics that we use to quantify the network's performance. The first two metrics are the interloper fraction mean-squared error,
\begin{equation}
    {\rm MSE} = \frac{1}{n_{\rm test}} \sum_{i=1}^{n_{\rm test}} \left( f_i - y_{f, i} \right)^2 \, ,
    \label{eq:fMSE}
\end{equation}
which gives an estimate of the goodness of the network's predicted mean, and the reduced chi-square, 
\begin{equation}
    \chi_{\rm red}^2 = \frac{1}{n_{\rm test}} \sum_{i=1}^{n_{\rm test}} \frac{\left( f_i - y_{f, i} \right)^2 }{\sigma_{f, i}^2} \, , 
    \label{eq:fchi2}
\end{equation}
 that quantifies the accuracy of the estimated error. For the correction, we define a metric for the first moment, which is the mean correction error,
\begin{equation}
    {\rm MCE}(k) = 1 - \frac{1}{n_f} \sum_{i=1}^{n_f} \frac{y_{{\rm c}, i}(k)}{\widehat{y}_{{\rm c}, i}(k)} \, ,
    \label{eq:MCE}
\end{equation}
where the summation is over the simulations with interloper fraction within a given range ($f_{\rm min} < f < f_{\rm max}$) and $n_f$ is the number of simulations in this range. This metric quantifies the residual error in the power spectrum monopole after the network correction. In general, we consider the correction good if its error is below $1\%$ in all $k$-bins. Then we compute the error of the MCE, starting from the second moment of the power spectrum correction outputted by the network, through error propagation and dividing it by the square root of the number of simulations. This gives us an estimate of the bias in the network correction.

%%%%%%%%%%%%%%%%%%%%%%%%%%%%%%%%%%%%%
\section{Results}
\label{sec:results}
%%%%%%%%%%%%%%%%%%%%%%%%%%%%%%%%%%%%%

We now present the performance of the method across the fixed and varying cosmology setups.

%-------------------------------
\subsection{Fixed Cosmology}
\label{sec:fixed-cosmo}
%-------------------------------
%
\begin{figure*}\centering
\includegraphics[width=.45\textwidth]{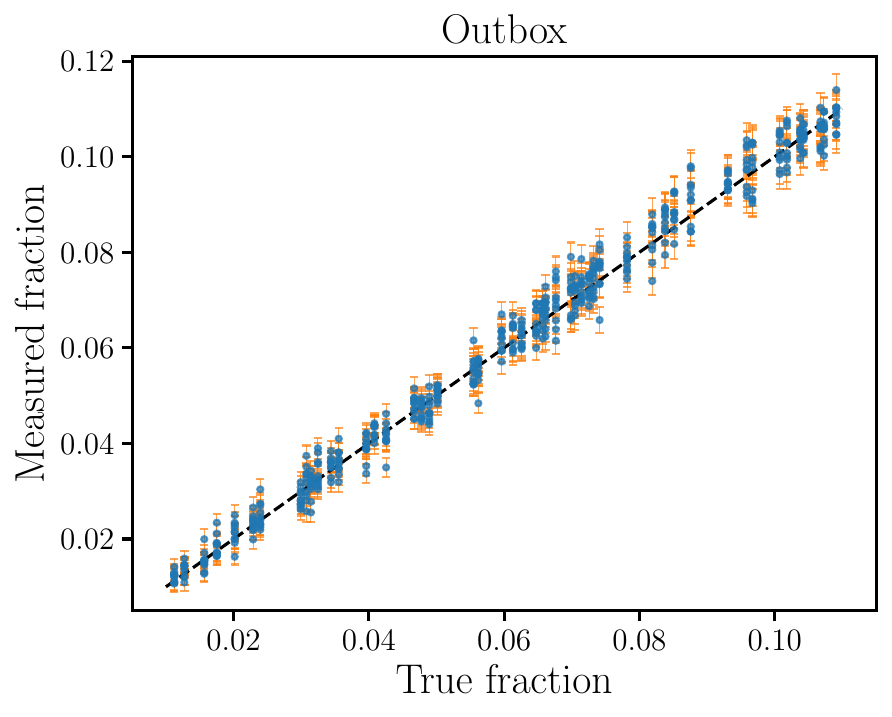}
\includegraphics[width=.45\textwidth]{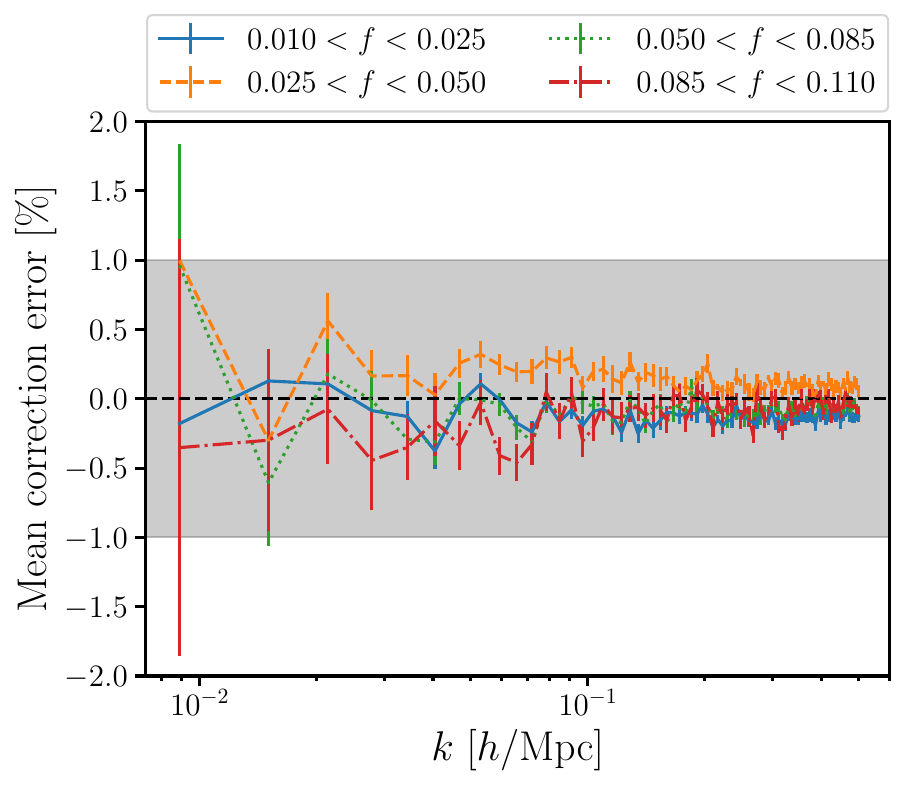}
\includegraphics[width=.45\textwidth]{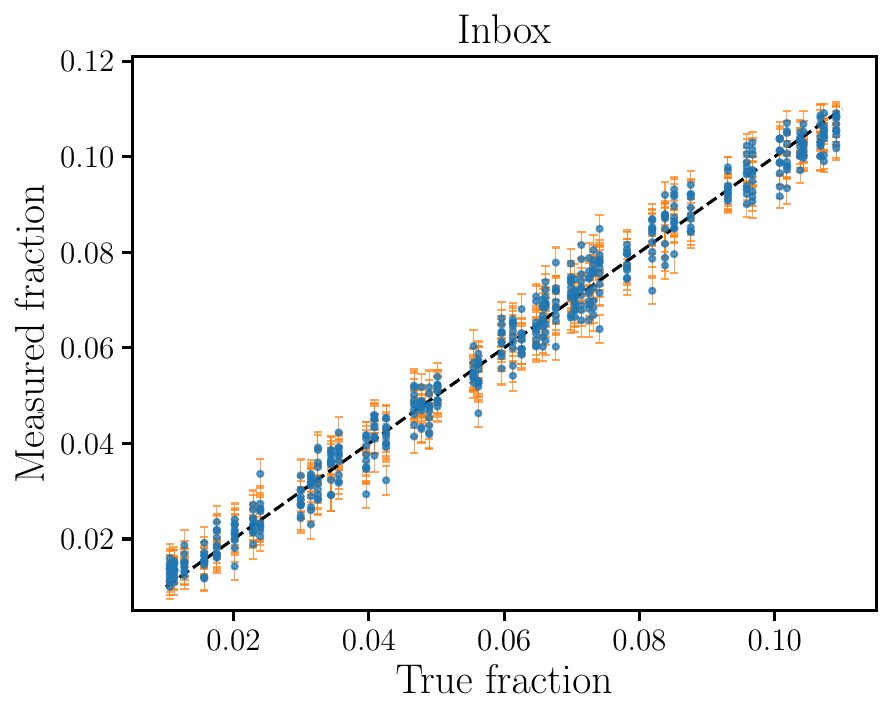}
\includegraphics[width=.45\textwidth]{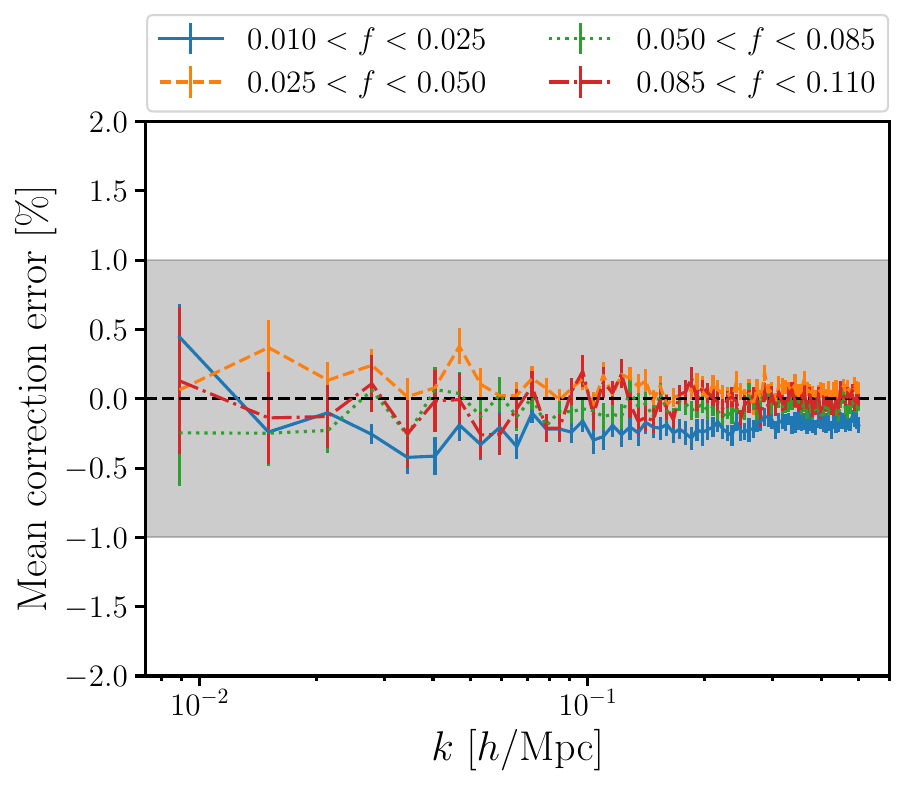}%
\caption{Results in fixed cosmology for the outbox (top) and inbox (bottom) interlopers when the network input is the contaminated power spectrum monopole. \emph{Left}: True versus inferred interloper fraction. \emph{Right}: Mean power spectrum monopole correction error. The shaded area indicates a $1\%$ error threshold. Error bars represent the mean of the predicted uncertainties within each bin, divided by the square root of the number of objects in the corresponding fraction range.}\vspace{0.1in}
\label{fig:outbox-inbox-fixed} 
\end{figure*}

As discussed in section~\ref{sec:nointernet}, the network learns to infer both the fraction of interlopers and the correction on the power spectrum monopole from either the contaminated power spectrum monopole or the combination of the power spectrum and bispectrum monopoles. The first two metrics we use to evaluate the network performance are the fraction MSE and $\chi_{\rm red}^2$, as defined in section~\ref{sec:proc-metrics}. Although the primary goal of the network is to correct the power spectrum monopole, it also returns an estimate of the interloper fraction. We observe a strong correlation between the accuracy of this estimate, quantified by the mean squared error, and the quality of the resulting power spectrum correction. For this reason, we use fraction-based metrics as an initial proxy for evaluating the algorithm's overall performance.

In Table~\ref{tab:metrics-fixed}, we report the MSE and $\chi_{\rm red}^2$ of the inferred fractions. When considering the MSE, the inclusion of bispectrum information leads to improved performance for both interloper types, yielding an improvement of approximately $30\%$ for the outbox case and around $4\%$ for the inbox interlopers. However, the improvement in the MSE corresponds to a degradation of the $\chi_{\rm red}^2$ in both cases. In particular, for the inbox interlopers, the $\chi_{\rm red}^2$ drastically increases due to a larger scatter in the fraction predictions that does not correspond to larger predicted errors. Nevertheless, in all cases, the $\chi_{\rm red}^2$ is larger than $1$, indicating that the network tends to underestimate the error on the fraction. Given the better $\chi_{\rm red}^2$ values of the network that takes as input the contaminated power spectrum, in the next two paragraphs we concentrate on the results of this configuration.

The left panels in figure~\ref{fig:outbox-inbox-fixed} show scatter plots comparing the true and predicted interloper fractions, along with the corresponding inferred uncertainties, for the outbox and inbox cases. Indeed, we observe that the predicted fractions are tightly centered around their true values, even near the edges of the prior, which corresponds to a low MSE. However, the associated uncertainties appear overconfident, as several points lie multiple $\sigma$ away from their true values, resulting in $\chi_{\rm red}^2 > 1$.

Finally, the right panels of figure~\ref{fig:outbox-inbox-fixed} show the mean correction error for different values of interloper fraction. The shaded area indicates a $1\%$ error margin. We find that, across all fraction ranges, the correction remains well within this threshold for both the outbox and inbox cases. In the outbox case, the mean correction exhibits virtually no bias across all fraction ranges, while increased scatter is observed at low $k$, with broader error bars in these bins reflecting the expected variability. For the inbox interlopers, the network appears slightly overconfident across all scales, yielding mildly biased results, though these remain well within the $1\%$ error margin. The lowest fraction range, $0.001 < f < 0.025$ (blue solid line), exhibits a more noticeable bias. This is attributed to the network's tendency to overestimate the fraction, which leads to an over-correction of the power spectrum monopole and results in a negative mean correction (see equation~\ref{eq:MCE}).

Overall, the network performs slightly better for outbox interlopers, both in terms of fraction estimation metrics and correction of the power spectrum monopole. Nevertheless, for both interloper types, it achieves low fraction MSE values and effective monopole correction. 

\begin{table*}
\caption{\label{tab:metrics-BSQ}Fraction MSE and $\chi_{\rm red}^2$ values when varying cosmology. The $\chi_{\rm red}^2$ values marked by * are computed removing the points with predicted uncertainty $<10^{-3}$, which are always less than $2\%$ of the test set.}
\begin{ruledtabular}
\begin{tabular}{cccccc}
Input & Priors & \multicolumn{2}{c}{Outbox} & \multicolumn{2}{c}{Inbox} \\ \midrule
 & & MSE & $\chi_{\rm red}^2$ & MSE & $\chi_{\rm red}^2$ \\ \midrule
\multirow{4}{*}{contam. $P_0$} & known cosmology & $ 5.34 \times 10^{-5}$ & 0.96* & $4.88 \times 10^{-5}$ & 1.33* \\ \cmidrule(l){2-6}
& \multirow{1}{*}{$1 \times \sigma_{\rm Planck}$} & $1.49 \times 10^{-4}$ & 1.33* & $1.05 \times 10^{-4}$ & 1.08* \\ \cmidrule(l){2-6}
& \multirow{1}{*}{$3 \times \sigma_{\rm Planck}$} & $2.77 \times 10^{-4}$ & 0.98 & $1.75 \times 10^{-4}$ & 1.29* \\ \cmidrule(l){2-6}
& \multirow{1}{*}{$5 \times \sigma_{\rm Planck}$} & $3.47 \times 10^{-4}$ & 1.11 & $2.29 \times 10^{-4}$ & 1.66* \\ \midrule
\multirow{4}{*}{contam. $P_0 + B_0$} & known cosmology & $2.62 \times 10^{-5}$ & 1.58* & $3.77 \times 10^{-5}$ & 1.22 \\ \cmidrule(l){2-6}
& \multirow{1}{*}{$1 \times \sigma_{\rm Planck}$} & $9.29 \times 10^{-5}$ & 1.25* & $8.17 \times 10^{-5}$ & 1.33 \\ \cmidrule(l){2-6}
& \multirow{1}{*}{$3 \times \sigma_{\rm Planck}$} & $1.36 \times 10^{-4}$ & 1.16* & $1.26 \times 10^{-4}$ & 1.38* \\ \cmidrule(l){2-6}
& \multirow{1}{*}{$5 \times \sigma_{\rm Planck}$} & $1.52 \times 10^{-4}$ & 1.39* & $1.58 \times 10^{-4}$ & 1.33*
\end{tabular}
\end{ruledtabular}\vspace{0.1in}
\end{table*}

Compared to \citet{Massara:2023ull}, our method yields almost an order of magnitude improvement on the MSE (see their figure~2), albeit with a $\chi^2_{\rm red}$ $6\%$ larger. As the halo density of the two samples is comparable ($n \sim 2 \times 10^{-4} \, h^3/{\rm Mpc}^3$), we attribute this improvement in performance to the larger volume we utilize. Indeed, in our work we use the whole \textsc{Quijote} boxes, which have a volume of $1 \, {\rm Gpc}^3/h^3$, while \cite{Massara:2023ull} analyses cropped boxes with volume $0.0225 \, {\rm Gpc}^3/h^3$.

%-------------------------------
\subsection{Varying Cosmology}
\label{sec:vary-cosmo}
%-------------------------------

After testing the performance of the network when cosmology is fixed, we consider the more realistic case where cosmological parameters are varied together with the interloper fraction. We use the big Sobol sequence \textsc{Quijote} simulation suite (see section~\ref{sec:quijote}) to generate the training data. In this configuration, in addition to using the contaminated power spectrum monopole or its combination with the bispectrum monopoles as inputs, we also provide the network with priors on the cosmological parameters. Specifically, we input parameter values sampled from a Gaussian distribution centered on the true simulation values, with a standard deviation scaled by a multiple of the Planck posteriors \citep{2020A&A...641A...9P}.

In this configuration, we trained the network for both interloper types using different prior widths, considering either the power spectrum alone or its combination with bispectrum information.

%~~~~~~~~~~~~~~~~~~~~~~~~~~~~~
\subsubsection{Correcting with Power Spectrum Information}
\label{sec:vc-p0p0}
%~~~~~~~~~~~~~~~~~~~~~~~~~~~~~

\begin{figure*}[htbp!]
\centering
\includegraphics[width=.45\textwidth]{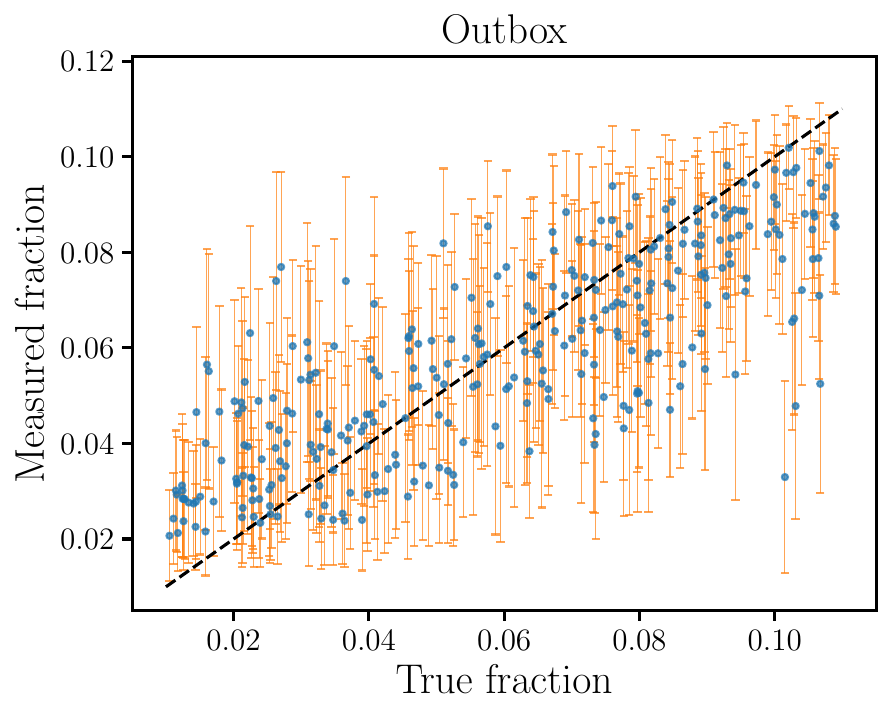}\includegraphics[width=.45\textwidth]{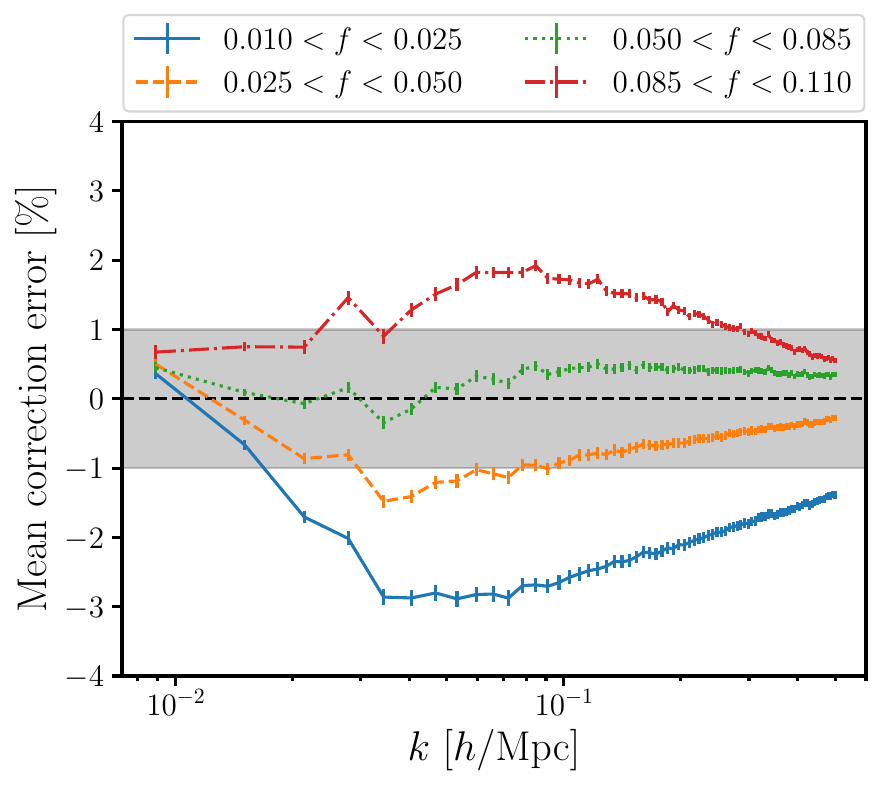}
\includegraphics[width=.45\textwidth]{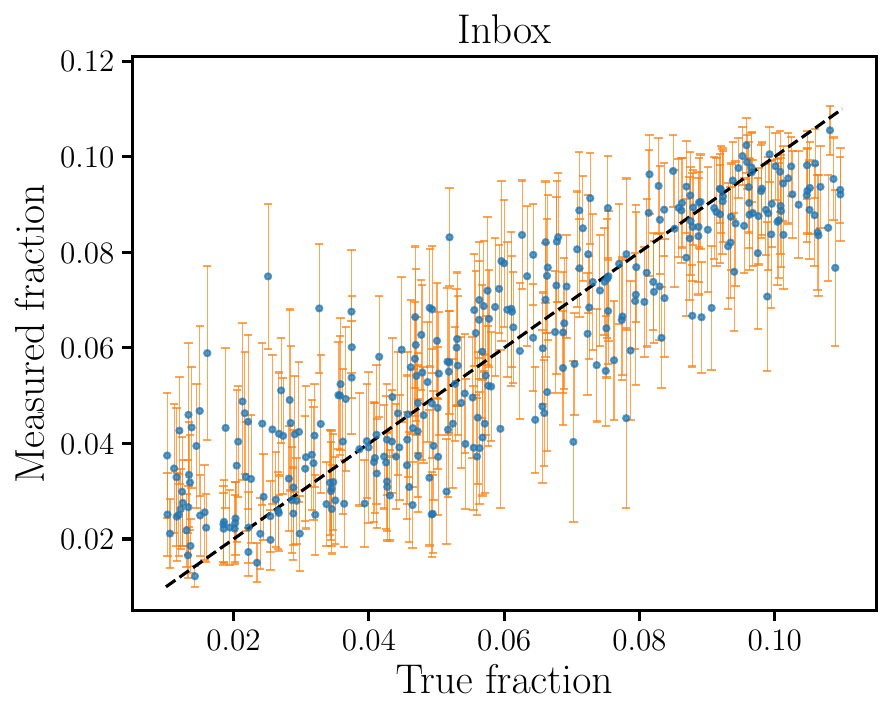}\includegraphics[width=.45\textwidth]{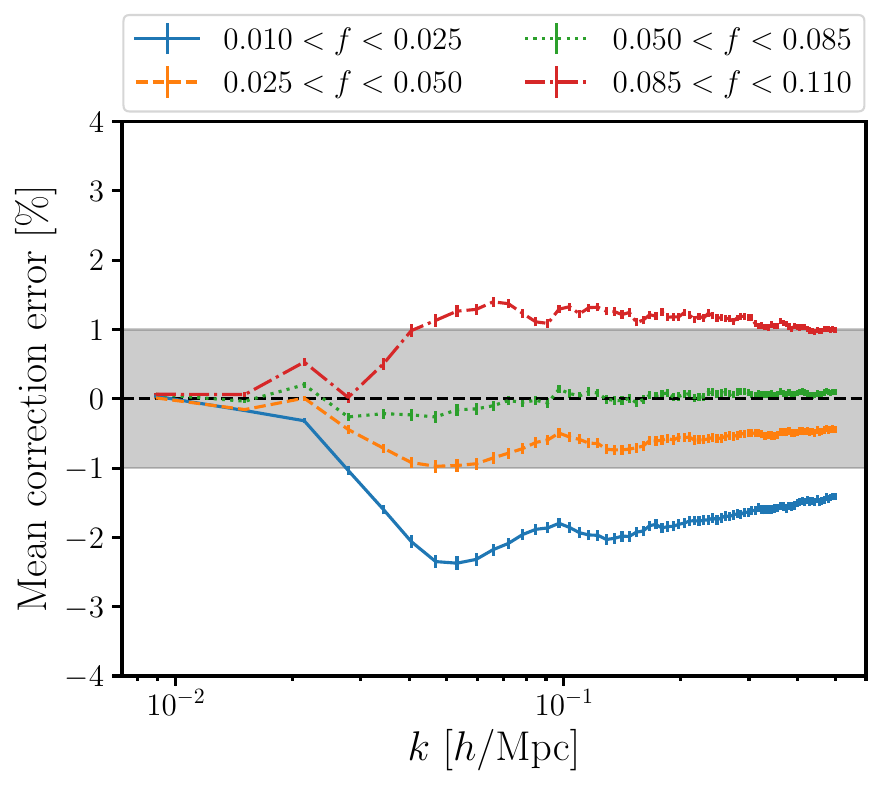}%
\caption{Results for varying cosmology with $3 \times \sigma_{\rm Planck}$ priors and contaminated power spectrum monopole as input for the outbox (top) and inbox (bottom) interlopers. \emph{Left}: Scatter plot of true versus measured fraction for a subsample of the test set. \emph{Right}: Mean correction error on the power spectrum monopole for four true fraction ranges. The shaded area indicates the $1\%$ error range.}
\label{fig:outbox-inbox-P0-BSQ} \vspace{0.1in}
\end{figure*}

Table~\ref{tab:metrics-BSQ} reports the MSE and $\chi_{\rm red}^2$ for outbox and inbox interlopers, considering three prior widths on cosmological parameters ($1\times$, $3\times$, and $5 \times \sigma_{\rm Planck}$ uncertainties). Notably, the network does not achieve the same MSE performance as in the fixed cosmology case, even when provided with the fiducial cosmological parameters from the simulations (referred to as `known cosmology'). However, the $\chi_{\rm red}^2$ values are consistently closer to $1$ in the varying cosmology setups, indicating that the network appropriately accounts for the increased uncertainties due to variation of cosmology.

As expected, broader priors degrade the accuracy of the network's predictions, leading to nearly an order-of-magnitude increase in MSE between the best and worst scenarios for both interloper types. Nevertheless, the $\chi_{\rm red}^2$ remains stable and close to $1$, further demonstrating the network’s ability to accommodate the added uncertainty from variation of cosmology.

We note that, in computing $\chi_{\rm red}^2$, we excluded simulations from the test set in which the predicted uncertainty was below $10^{-3}$, as such predictions are deemed unreliable. Typically, this excluded subset comprises fewer than $10$ simulations out of a sample of over $3000$, and in rare cases up to about $50$, but always fewer than $100$.

Figure~\ref{fig:outbox-inbox-P0-BSQ} presents scatter plots comparing the true and predicted interloper fractions (left panels) and the mean correction error on the power spectrum monopole (right panels) for the outbox (top) and inbox (bottom) interloper cases, using networks trained with $3 \times \sigma_{\rm Planck}$ cosmological priors. Unlike in the fixed cosmology case, the network now performs better in correcting the power spectrum for inbox interlopers than for outbox interlopers. A possible explanation for this behavior is that, in the inbox case, the network exploits the BAO shift to disentangle the cosmology and interloper effects. However, we would require further testing to confirm this. For both interloper types, the fraction scatter plots reveal deviations near the edges of the training range, which correspond to poorer corrections in the power spectrum monopole, particularly in the ranges $0.001 < f < 0.025$ (solid blue line) and $0.085 < f < 0.110$ (dash-dotted red line). Specifically, in the lowest fraction range, the network tends to overestimate the fraction, resulting in an over-correction of the power spectrum. Conversely, for the highest fractions, it underestimates the fraction, leading to under-correction.

It is worth noting that, in contrast to the fixed cosmology case, where the network successfully corrects the power spectrum monopole across all scales, the correction exhibits a scale dependence when cosmology is varied. This trend can be attributed to the network's difficulty in disentangling the effects of varying interloper fraction and cosmological parameters on the power spectrum monopole. The scale dependence of the error on the correction differs in the inbox and outbox case, with the latter clearly showing oscillatory variations corresponding to BAO and a flatter behavior at larger $k$s. We note that the leftover error shows a peak that is related to the scale of the interloper shift.

In this configuration, a comparison with \cite{Massara:2023ull} is less straightforward, as they vary cosmology, interloper fraction, and halo mass cut to simulate different biases (see their figure~4). If we compare their best performing configuration with a fixed density of $n = 7.2 \times 10^{-5} \, h^3/{\rm Mpc}^3$ in boxes of $0.0625 \, {\rm Gpc}^3/h^3$, and a $1 \times \sigma_{\rm Planck}$ prior, we perform about $4$ times better even with larger cosmology prior ($5 \times \sigma_{\rm Planck}$). However, we stress that our analysis setup has a larger volume, a denser halo sample, and we are not marginalizing over the halo bias.

%~~~~~~~~~~~~~~~~~~~~~~~~~~~~~
\subsubsection{Correcting with Power Spectrum and Bispectrum Information}
\label{sec:vc-p0b0p0}
%~~~~~~~~~~~~~~~~~~~~~~~~~~~~~

\begin{figure*}[htbp!]
\centering
\includegraphics[width=.45\textwidth]{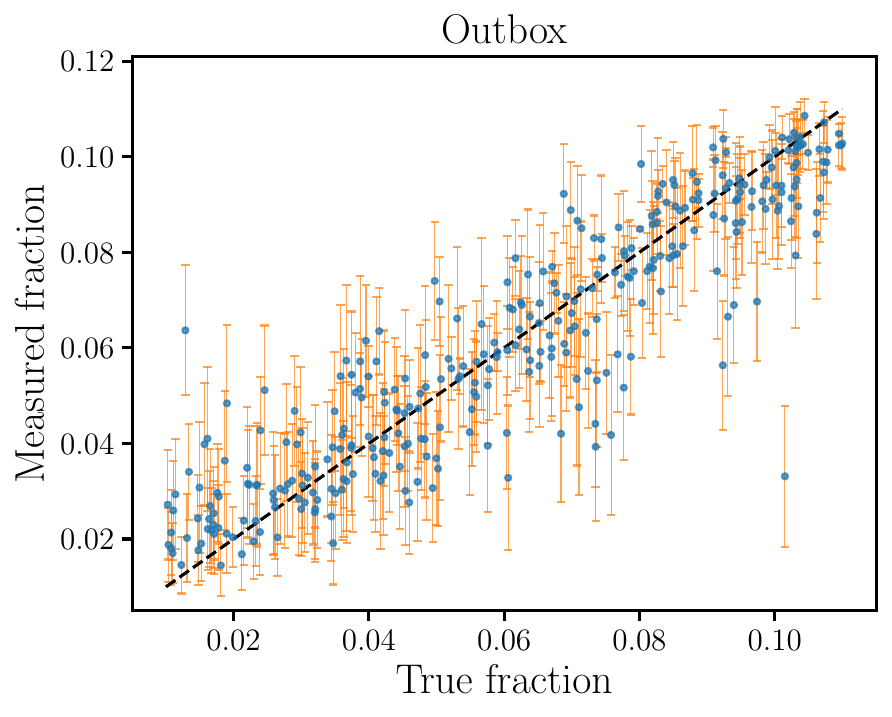}\includegraphics[width=.45\textwidth]{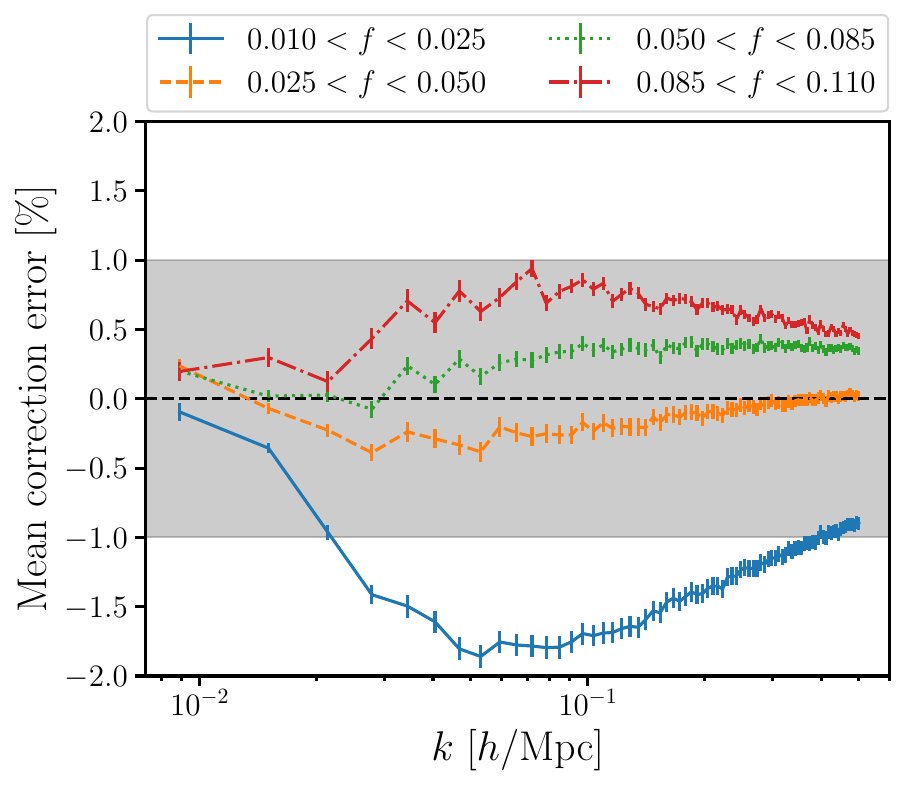}
\includegraphics[width=.45\textwidth]{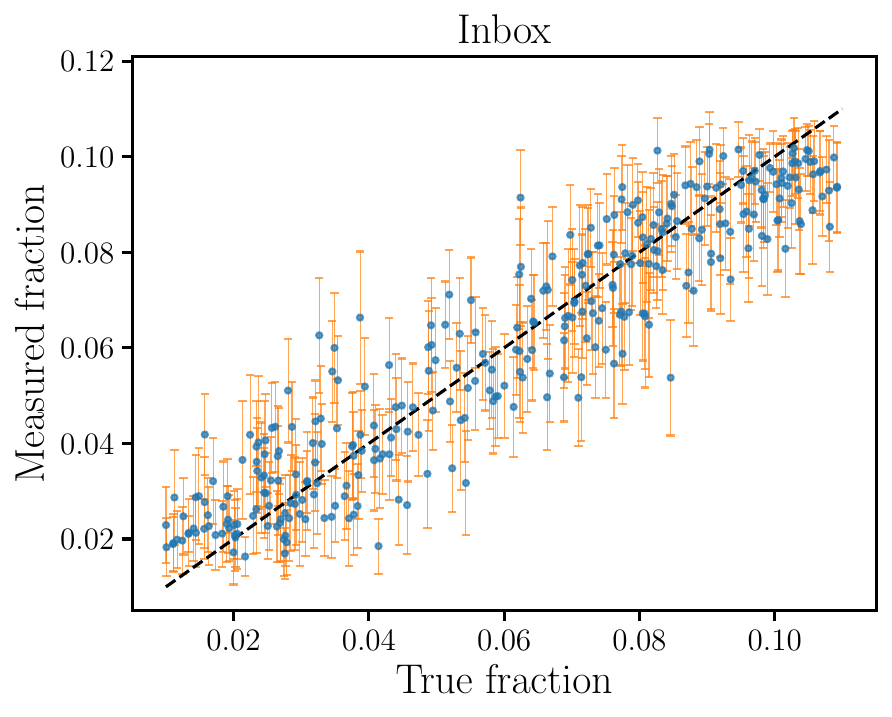}\includegraphics[width=.45\textwidth]{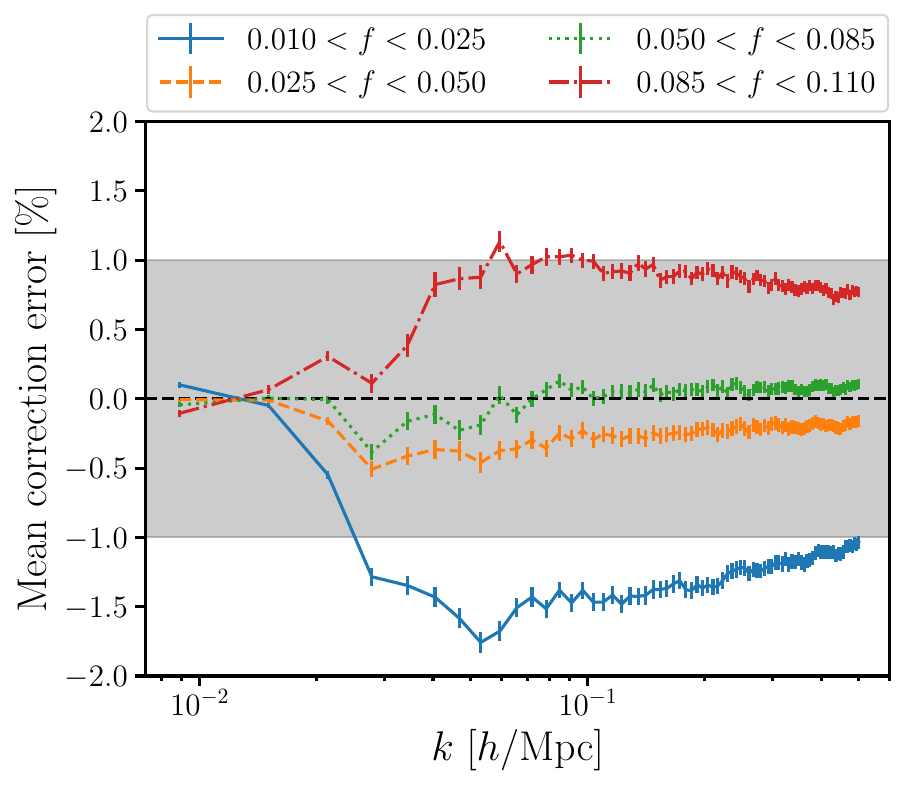}%
\caption{Same as figure~\ref{fig:outbox-inbox-P0-BSQ} but taking contaminated power spectrum and bispectrum monopoles as input. We note that there is a change in the $y$-axis range between this figure and the right panels of figure~\ref{fig:outbox-inbox-P0-BSQ}.}\vspace{0.1in}
\label{fig:outbox-inbox-P0B0-BSQ}
\end{figure*}

\begin{figure*}%[htbp!]
\includegraphics[width=.5\textwidth]{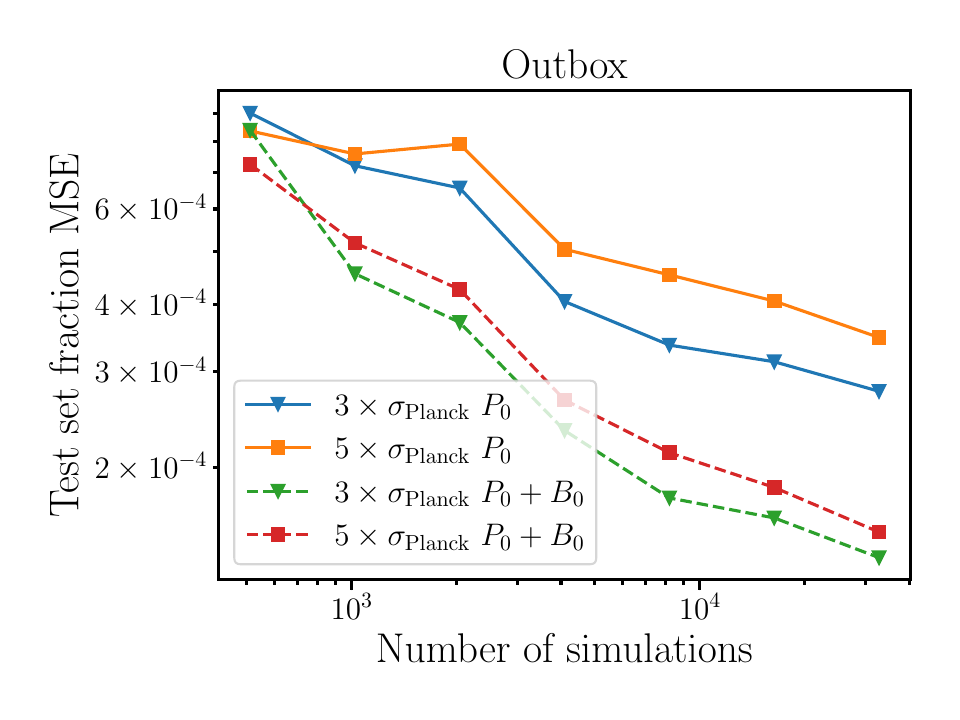}
\includegraphics[width=.5\textwidth]{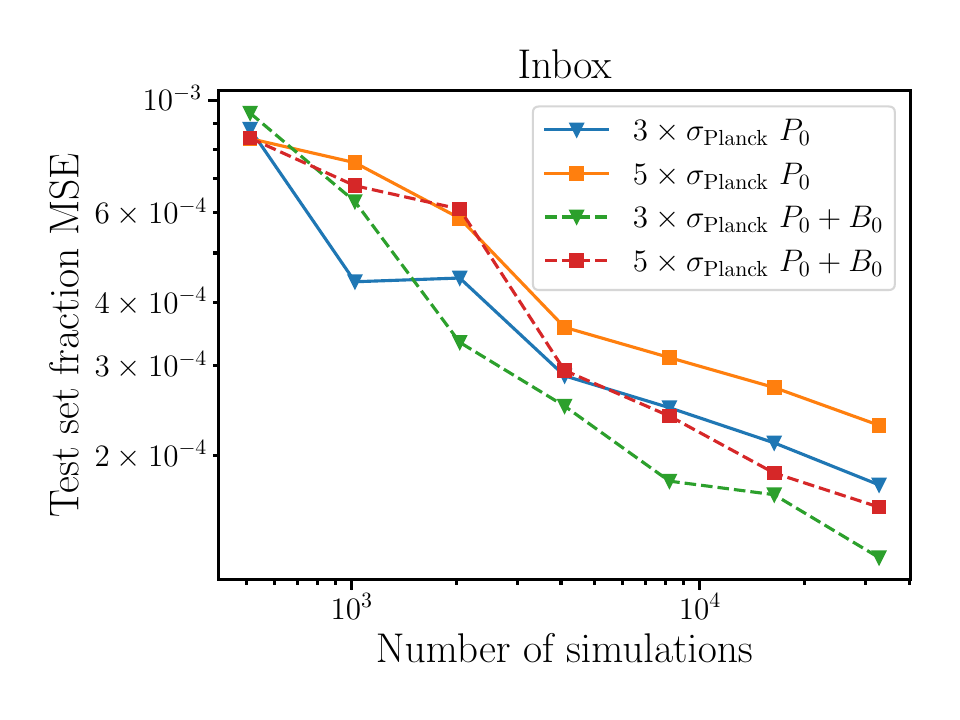}%
\caption{\label{fig:MSE-N} Mean squared error on the test set as a function of the number of simulations in the whole dataset. The reverse triangular markers identify the networks trained with $3 \times \sigma_{\rm Planck}$ priors on the cosmological parameters, the squares the $5 \times \sigma_{\rm Planck}$ priors. The solid lines refer to the results from the networks that take as input the contaminated power spectrum monopole and the dashed line to the network with contaminated power spectrum and bispectrum monopoles as inputs. \emph{Left}: results for the outbox interlopers. \emph{Right}: results for the inbox interlopers.}\vspace{0.1in}
\end{figure*}

Next, we trained the network to jointly predict the interloper fraction and the cleaned power spectrum monopole, using the contaminated power spectrum and bispectrum monopoles combined with cosmological priors as input. We find that, under varying cosmological parameters, the inclusion of the bispectrum monopole significantly enhances network performance for both outbox and inbox interlopers. For outbox interlopers, the fraction MSE improves by approximately $40$--$60\%$, while for inbox interlopers the improvement is slightly less, ranging from $20$--$30\%$. However, the $\chi_{\rm red}^2$ values are generally larger than those from the network using only the contaminated power spectrum monopole as input, and consistently exceed $1$, indicating that the network is somewhat overconfident in its uncertainty estimates.

The fraction scatter plots in the left panels of figure~\ref{fig:outbox-inbox-P0B0-BSQ} show a noticeably tighter distribution, although edge effects near the bounds of the fraction prior persist. This is particularly evident at the lower end of the fraction range, where the mean correction error in the power spectrum monopole is larger (see solid blue line in the right panels of the same figures). The second most significant bias occurs at the high-fraction end (dash-dotted red line), though in this case the mean correction remains within the $1\%$ error band. Additionally, when the MSE gets below the $1\%$ error, the correction becomes more efficient at all scales, reducing the scale dependence of the residual error. For inbox interlopers, while a slight increase in correction error persists at small scales, the distinct oscillatory feature observed earlier is significantly suppressed. Finally, we stress again the fact that the larger residual error we observe for the low and high interloper fractions is related to the network hitting the training prior edges. Therefore, for survey-specific applications, we will have to center the interloper fraction prior on the expected value to minimize this effect.

We also note the presence of a few outliers in the fraction scatter plots, which lead to catastrophic errors in the power spectrum correction. Upon inspection, these outliers correspond to simulations with extreme cosmological parameter values and are therefore not expected to reflect the network's behavior when applied to real data. 

%~~~~~~~~~~~~~~~~~~~~~~~~~~~~~
\subsubsection{Scaling with  Number of Simulations}
\label{sec:Nsimulations}
%~~~~~~~~~~~~~~~~~~~~~~~~~~~~~

The results presented in the previous two subsections were obtained using the full \textsc{Quijote} BSQ suite. We now assess the network's \emph{scaling law}, which is its performance as a function of the number of simulations used in the analysis. In first approximation, scaling laws have three main regimes: the small data region, where the network error slowly decreases due to the reduced amount of data in the training set, the power-law region, where the error very efficiently decreases with the number of training data, and the irreducible error region, where the network's error plateaus \citep{2017arXiv171200409H,2020arXiv200108361K,2024PNAS..12111878B}. By determining the scaling law of the network, we understand if we have reached the learning plateau, or if there is still room for improvement by increasing the size of the training set.

For this analysis, we construct progressively larger subsets of the Sobol sequence, doubling the dataset size at each step, from $512$ simulations up to the full set of $2^{15}$ simulations. Each subset is divided into training ($75\%$), validation ($15\%$), and test ($10\%$) sets.

Figure~\ref{fig:MSE-N} shows the MSE on the test set as a function of the total number of simulations, for two prior widths on cosmological parameters ($3 \times$ and $5 \times \sigma_{\rm Planck}$, represented by reverse triangles and squares), two types of network input (the contaminated power spectrum monopole alone, or combined with the bispectrum monopole, shown as solid and dashed lines), and the two interloper types (outbox and inbox in the left and right panels). We notice that in both cases we have not yet reached the plateau of the scaling law, showing that the network's estimates have not converged yet.

In the outbox case, the slopes of the MSE curves differ significantly depending on whether the bispectrum information is included. Regardless of prior width, incorporating the bispectrum monopole consistently improves performance compared to using the power spectrum alone, achieving comparable or better results with fewer simulations.

For inbox interlopers, a different trend emerges. The MSE curves exhibit similar slopes, and the impact of the input choice (power spectrum versus power spectrum plus bispectrum) is less pronounced. Still, the best-performing configuration corresponds to the network using both the power spectrum and bispectrum monopoles with the tighter cosmological priors (reverse triangles with dashed green line), while the worst performance arises from the network using only the power spectrum with wider priors (squares with solid orange line). Notably, the other two configurations yield comparable results, and the network using only the power spectrum monopole with tighter priors outperforms the one using bispectrum information with wider priors, highlighting the importance of prior constraints.

%%%%%%%%%%%%%%%%%%%%%%%%
\section{Conclusions}
\label{sec:conclusions}
%%%%%%%%%%%%%%%%%%%%%%%%

In this work, we presented a proof-of-concept study demonstrating the use of machine learning to correct for line interlopers at the level of the measured summary statistics. Using the \textsc{Quijote} simulations, we constructed snapshots with interlopers coming from redshifts similar to the target sample, inbox interlopers, or from very different redshifts, outbox interlopers, which are available within the \textsc{Quijote} suite. We measured the interloper-contaminated and the target summary statistics (power spectrum and bispectrum) in redshift space and trained a neural network to infer the correction to the power spectrum monopole and the interloper fraction from the contaminated statistics. We have focused on cleaning the power spectrum monopole only since the measurements of the quadrupole and hexadecapole in the \textsc{Quijote} simulations were too noisy due to the limited box size.

We evaluated the network's performance separately for outbox and inbox interlopers under various configurations. First, we used the contaminated power spectrum monopole as input, and then tested whether including the bispectrum monopole improved the results. Additionally, we trained the network under two scenarios: one where cosmology was fixed and only the interloper fraction varied, and another where both the cosmological parameters and the interloper fraction were allowed to vary. In the latter case, we also provided the network with prior information on cosmology.

We found that, when cosmology is fixed, the network accurately infers the interloper fraction and effectively corrects the power spectrum monopole for inbox and outbox interlopers alike. In particular, the residual error in the power spectrum is lower than $1\%$. 

In contrast, when both cosmology and the interloper fraction are allowed to vary, we observed a degradation in the network's performance. In this configuration, prior edge effects become apparent near the boundaries of the fraction range ($f \in [0.01, 0.11]$), leading to biased corrections. We emphasize that these distortions arise from the network encountering the limits of the prior, which implies that, given an expected interloper fraction for a specific experiment, this issue can be mitigated by appropriately adjusting the prior range. Additionally, we observed a significantly larger scatter in the predicted interloper fractions. This effect is present even when the network is provided with the true cosmological parameters of the simulations and becomes more pronounced as the width of the cosmological priors increases.

Additionally, in the scenario where both cosmology and the interloper fraction vary, we demonstrated that incorporating the bispectrum monopole significantly enhances the network’s ability to correct the power spectrum monopole. Specifically, it reduces the interloper fraction MSE by $40$--$60\%$ for outbox interlopers and by $20$--$30\%$ for inbox interlopers. However, this improvement comes at the cost of increased $\chi^2_{\rm red}$ values for the fraction, indicating that the network tends to be overconfident in its uncertainty estimates. A similar, though less pronounced, trend is observed when the bispectrum monopole is added in the fixed cosmology setup. We also emphasize that the few catastrophic correction errors encountered in the varying cosmology case correspond to simulations with extreme cosmological parameter values, which are regions of parameter space unlikely to be encountered in real-data applications.

Finally, in the varying cosmology scenario, we also evaluated the network's performance as a function of the number of simulations used for training. Interestingly, we observed distinct behaviors for outbox and inbox interlopers. For outbox interlopers, the inclusion of bispectrum information yielded the most significant improvements, whereas tightening the cosmological priors had a comparatively smaller impact. This suggests that, to achieve a given target MSE for the interloper fraction, similar or better performance can be obtained with fewer simulations by incorporating bispectrum information rather than relying on stricter priors. In contrast, this trend does not hold for inbox interlopers. When the number of available simulations is limited, tightening the cosmological priors appears to be more beneficial than adding bispectrum information. This difference in behavior may stem from the varying levels of correlation between the interlopers and the target sample. Specifically, inbox interlopers are more strongly correlated with the target galaxies and have a notable impact on the BAO features in the power spectrum, making the learning process more sensitive to cosmological parameter constraints.

The approach introduced in this study yields promising results in correcting the power spectrum monopole. However, before it can be applied to real data, the method must be tested under more complex and realistic conditions. For example, upcoming surveys like Euclid are expected to face contamination from multiple types of line interlopers, such as [S${\rm III}$] from low redshift and [O${\rm III}$] from high redshift, as well as from noise interlopers. A natural extension of this work would be to introduce multiple line interlopers simultaneously in the simulations and assess whether the network can disentangle their individual contributions while still accurately correcting the power spectrum. It would also be important to test the method’s ability to handle noise interlopers, which are stochastic in nature. 

To address the challenge of correcting for multiple types of interlopers, it may be beneficial to work in configuration space. In the observed two-point correlation function, the BAO peak results from the superposition of contributions from both the target galaxies and the interlopers. Interloper BAO peaks are typically shifted to smaller or larger scales relative to that of the target population. Consequently, the observed BAO feature can take on a distinctive shape, potentially encoding information about contamination fractions and helping to inform corrections in a machine learning–based analysis.

Two further extensions include applying the algorithm to lightcone simulations, incorporating survey-specific geometries and the power spectrum quadrupole into the correction process. As noted earlier, we did not include the quadrupole in this study as its measurements in the \textsc{Quijote} boxes are extremely noisy. Nevertheless, the quadrupole is expected to contain additional information about interloper contamination that could improve monopole correction. Moreover, because the quadrupole is commonly used in modern cosmological analyses, the ability to remove interloper contamination from it would be particularly beneficial.

\section*{Acknowledgments}
We thank Enzo Branchini and Ilaria Risso for insightful discussions and comments on the first version of the draft. The simulations and analysis of this work have been done thanks to the facilities offered by the Univ. Savoie Mont Blanc - CNRS/IN2P3 MUST computing center.

%\appendix

\bibliography{main}{}
\bibliographystyle{aasjournal}

%% This command is needed to show the entire author+affiliation list when
%% the collaboration and author truncation commands are used.  It has to
%% go at the end of the manuscript.
%\allauthors

%% Include this line if you are using the \added, \replaced, \deleted
%% commands to see a summary list of all changes at the end of the article.
%\listofchanges

\end{document}